\documentclass[11pt,a4paper]{article}

\usepackage[T1]{fontenc}
\usepackage[utf8]{inputenc}
\DeclareUnicodeCharacter{2010}{-}
\usepackage{lmodern}
\usepackage[margin=1in]{geometry}
\usepackage{amsmath,amssymb,amsthm}
\usepackage{graphicx}
\usepackage{bm}
\usepackage{indentfirst}
\usepackage{comment}
\usepackage{verbatim}
\usepackage{enumitem}
\usepackage{float}
\usepackage{algorithm}
\usepackage{algpseudocode}
\usepackage{caption}
\usepackage[numbers,square,sort&compress]{natbib}
\usepackage[hidelinks]{hyperref}

\floatstyle{ruled}
\restylefloat{algorithm}
\DeclareCaptionLabelFormat{namenumberless}{#1}
\theoremstyle{plain}
\newtheorem{theorem}{Theorem}

\newtheorem{lemma}{Lemma}
\newtheorem{observation}{Observation}

\theoremstyle{definition}

\theoremstyle{remark}

\title{An Optimal Online Algorithm for Single-machine Scheduling with Two Agents}

\author{%
  Peihai Liu\quad Kejun Zhao\quad Xiwen Lu\thanks{%
    Corresponding author.\\
    E-mail: \texttt{pliu@ecust.edu.cn} (P. Liu);
    \texttt{zhaokejun0711@163.com} (K. Zhao);
    \texttt{xwlu@ecust.edu.cn} (X. Lu)}\\[0.6em]
  {\normalsize School of Mathematics, East China University of Science and Technology,}\\
  {\normalsize Shanghai 200237, China}%
}
\date{}

\hypersetup{
  pdftitle={An Optimal Online Algorithm for Single-machine Scheduling with Two Agents},
  pdfauthor={Peihai Liu, Kejun Zhao, Xiwen Lu}
}

\begin{document}
\renewcommand{\qedsymbol}{}
\maketitle

\begin{abstract}
This paper addresses the online two-agent scheduling problem on a single machine, where jobs arrive over time. The objective is to minimize the weighted sum of the agents' makespans, denoted as $C_{\max}^A + \theta C_{\max}^B$, for a given priority weight $\theta \ge 1$. A significant gap has persisted between the established theoretical lower bound on the competitive ratio and the performance of the best-known online algorithms. 
For the  parameter range $1 \leq \theta \leq \tfrac{1+\sqrt{5}}{2}$, we propose novel online algorithm which achieves a competitive ratio of $1+\tfrac{\theta}{1+\theta}$. 
This result matches the theoretical lower bound.


\end{abstract}

\section{Introduction}\label{sec1}
The multi-agent scheduling problem is a modern extension of classical scheduling theory, capturing scenarios where multiple self-interested entities compete for shared resources. Originating from seminal works by Baker and Smith \cite{Baker03} and Agnetis et al. \cite{Agnetis04}, this domain addresses the challenge of coordinating jobs owned by different agents on common machines. Unlike traditional single-objective problems, multi-agent scheduling requires constructing schedules that balance the potentially conflicting performance objectives of all agents. The growing relevance of multi-agent scheduling is driven by practical demands in distributed systems, flexible manufacturing, and cloud computing, where the efficient coordination of shared resources across multiple users or clients is essential.
The increasing complexity of such systems has elevated multi-agent scheduling from a specialized topic to an important area of research, as documented by Agnetis et al. \cite{Agnetis14}.

The early works of Baker and Smith \cite{Baker03} and Agnetis et al. \cite{Agnetis04}, in the early 2000s, formally introduced two-agent scheduling problems, laying the groundwork for a vibrant and expanding research field. Since then, numerous models have been developed to capture the intricate interactions between agents. Broadly speaking, as described by Agnetis et al.\cite{Agnetis04},
there are three main classes of models: {\bf Constrained model}, {\bf Linear combination model} and {\bf Pareto optimization model}.

Over the years, these models have been enriched to incorporate increasingly realistic features. Variants with release dates \cite{Leung10}\cite{Yuan(2020)}, deadlines \citep{Wan(2015)}\cite{Ying2024}, and learning effects \citep{Shiau15} have been explored to better represent practical production environments. Further extensions include problems with rate-modifying activities \citep{Phosavanh2024,Phosavanh2025} and multi-objective formulations considering criteria such as minimizing late work and production cost \citep{Geng2025}. Collectively, these contributions demonstrate the continuous effort to bridge theoretical scheduling formulations with real-world applications.

This paper focuses on a problem within the \textbf{linear combination model}, namely the \textbf{online two-agent single-machine scheduling problem} with the objective of minimizing the weighted sum of makespans. In the online setting, jobs arrive sequentially over time, and scheduling decisions must be made immediately and irrevocably without any knowledge of future arrivals. This absence of future information introduces substantial complexity compared to the offline counterpart, which \citet{Ding2010} proved to be NP-hard. For the online version, \citet{Ding2010} established a competitive ratio lower bound of $1+\frac{\theta}{1+\theta}$ but proposed an algorithm achieving only a ratio of $2+\frac{1}{\theta}$, thereby creating a considerable gap between the best-known algorithm and the theoretical optimum. More recently, \citet{Lu2022} improved the upper bound to $\frac{1+2\theta+\theta^2}{2+2\theta+\theta^2}$; however, this result still does not attain the lower bound, leaving the development of an optimal online algorithm as an open challenge.

\noindent\textbf{Our contribution.}
In this paper, we resolve this long-standing open problem. We propose a novel online algorithm for the problem $1\mid r_j, \text{ online}\mid C_{\max}^A + \theta C_{\max}^B$, focusing on the challenging parameter range $1 \le \theta \le \frac{1+\sqrt{5}}{2}$. The proposed algorithm is designed to strategically balance the conflicting objectives of the two agents under the uncertainty inherent in the online environment. We provide a rigorous proof that our algorithm achieves a competitive ratio of $1+\frac{\theta}{1+\theta}$, thereby \textbf{matching the established lower bound}. This result closes the research gap identified by \citet{Ding2010} and improves upon the work of \citet{Lu2022}, providing a definitive and best possible solution for this class of online multi-agent scheduling problems.

The rest of the paper is organized as follows. Section~2 gives a formal statement of the problem and introduces notations. Section~3 presents our online algorithm for the case $1 \leq \theta \leq \tfrac{1+\sqrt{5}}{2}$. Section~4 provides a detailed analysis of its competitive ratio and proves its optimality. Finally, Section~5 concludes with a summary of our findings and discusses potential avenues for future research.

\section{Problem formulation and definitions}
\label{sec2}
\subsection{Problem formulation}

We consider a single-machine scheduling problem involving two competing agents, namely Agent A and Agent B. Let $\mathcal{J}^A$ and $\mathcal{J}^B$ denote the sets of jobs belonging to Agent A and Agent B, respectively. For simplicity, jobs in $\mathcal{J}^A$ and $\mathcal{J}^B$ are referred to as \textbf{A-jobs} and \textbf{B-jobs}. These two job sets are disjoint, i.e., $\mathcal{J}^A \cap \mathcal{J}^B = \emptyset$. Each job $J_j^X \in \mathcal{J}^X$ (for $X \in \{A, B\}$) is characterized by a processing time $p_j^X$ and a release date $r_j^X$. All jobs are to be processed non-preemptively on a single machine.

Each agent $X \in \{A, B\}$ seeks to minimize its makespan, defined as $C_{\max}^X = \max_{j \in \mathcal{J}^X} \{C_j^X\}$, where $C_j^X$ is the completion time of job $J_j^X$. The overall objective is to find a schedule that minimizes a weighted sum of the two agents' makespans. The objective function is formulated as:
$\min C_{\max}^A + \theta C_{\max}^B$,
where $\theta \ge 1$ is a constant representing the relative importance of Agent B's objective.

This problem is studied in an \textbf{online setting}, where the existence and characteristics of any job $J_j^X$ are unknown until its release date $r_j^X$. At any time $t$, an online algorithm can either schedule an available job (one that has been released but not yet processed) or leave the machine idle. Decisions are irrevocable. Following the standard three-field notation \cite{Graham1979}, the problem under study is denoted as  
 $1 \mid r_j,\ \text{online} \mid C_{\max}^A + \theta C_{\max}^B$.

 \subsection{Definitions and Notations}
 
 We begin by introducing the key notations used throughout the analysis. 
A \textbf{decision point} $\boldsymbol{t}$ is defined as any time at which the machine is idle and at least one job is available. 
A job is \textit{available} at time $t$ if it has been released ($r_j^X \le t$) but not yet processed.

For clarity, notations are often presented for Agent A only; analogous definitions apply to Agent B.

\begin{figure}[htbp]
\centering
\includegraphics[width=0.5\textwidth]{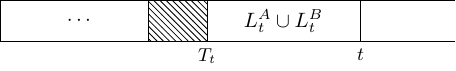}
\caption{An example of the notations defined at time $t$.}
\label{Fig1_sec2}
\end{figure}

\begin{itemize}
    \item $\bm{U_t^X}$: The set of available jobs for agent $X \in \{A, B\}$ at time $t$.
    \item $\bm{U_t}$: The set of all available jobs at time $t$, i.e., $\bm{U_t} = \bm{U_t^A} \cup \bm{U_t^B}$.
    \item $\bm{J_{s(t)}^A}$ and $\bm{J_{l(t)}^A}$: The available A-jobs in $\bm{U_t^A}$ with the shortest and longest processing times, respectively. Ties are broken arbitrarily.
    \item $\bm{T_t}$: The start time of the last contiguous block of jobs that was completed at time $t$. If the machine has been continuously busy since time 0, then $T_t = 0$.
    \item $\bm{L_t^A}$: The set of A-jobs that were processed within the time interval $[T_t, t]$.
    \item $\bm{P(S)}$: The total processing time of jobs in job set $S$, i.e., $P(S) = \sum_{J_j \in S} p_j$.
    \item $\bm{r_{\min}(S)}$: The minimum release date among jobs in job set $S$, i.e., $r_{\min}(S) = \min_{J_j \in S} \{r_j\}$.
\end{itemize}
Figure \ref{Fig1_sec2} provides a graphical illustration of these notations.

\section{An online algorithm}\label{sec3}
In this section, we propose a novel online algorithm for the considered problem. 
Our method, referred to as \textbf{AOSA} (Agent-Based Online Scheduling Algorithm), is designed to exploit the inherent trade-off between the makespans of the two agents. 
The algorithm dynamically adapts its scheduling decisions according to the composition of available jobs and the relative priority parameter~$\theta$.

When jobs from only one agent are available, AOSA postpones execution until a threshold time proportional to the job’s processing time and the weight~$\theta$ is reached. 
This delay prevents premature scheduling that could otherwise lead to an imbalance between the agents’ objectives.

When both agents have pending jobs, the algorithm compares their earliest release times and evaluates workload balance to decide whether to process a job from Agent~A or Agent~B. 
The decision rules are based on the cumulative processing requirements of both agents, adjusted by~$\theta$ and the start time of the current block. 
This mechanism ensures that machine time is allocated efficiently while respecting the relative priority encoded by~$\theta$.

Procedures~A and~B implement a selective strategy between the smallest and largest available jobs of the chosen agent. 
Scheduling the smallest job accelerates progress for the active agent, while scheduling the largest job prevents excessive postponement of long tasks that could dominate the makespan. 
The threshold inequalities in each procedure ensure that the choice aligns with minimizing the weighted objective $C_{\max}^A + \theta C_{\max}^B$.

\begin{algorithm}[H]
\renewcommand{\thealgorithm}{} 
\caption{AOSA (Agent-Based Online Scheduling Algorithm)}
\label{alg:AOSA}
\begin{algorithmic}[1]
\Require Priority weight $\theta \ge 1$; sequence of jobs arriving over time
\Ensure Feasible schedule minimizing $C_{\max}^A + \theta C_{\max}^B$

\While{there are unscheduled jobs}
    \State Determine $U_t$ (set of available unscheduled jobs at time $t$)
    \If{$U_t = U_t^A$}
        \If{$t \ge \frac{\theta}{1+\theta} \, p_{s(t)}^A$}
            \State Run \textbf{Procedure A}
        \Else
            \State Wait until $\frac{\theta}{1+\theta} \, p_{s(t)}^A$ or a new job arrives
        \EndIf
    \ElsIf{$U_t = U_t^B$}
        \If{$t \ge \frac{1}{1+\theta} \, p_{s(t)}^B$}
            \State Run \textbf{Procedure B}
        \Else
            \State Wait until $\frac{1}{1+\theta} \, p_{s(t)}^B$ or a new job arrives
        \EndIf
    \ElsIf{$U_t^A \neq \emptyset$ \textbf{and} $U_t^B \neq \emptyset$}
        \If{$r_{\min}(L_t^A \cup U_t^A) < r_{\min}(L_t^B \cup U_t^B)$}
            \If{$P(L_t^B \cup U_t^B) \ge \theta P(L_t^A \cup U_t^A) - (1+\theta)T_M$}
                \State Run \textbf{Procedure A}
            \Else
                \State Run \textbf{Procedure B}
            \EndIf
        \Else
            \If{$P(L_t^B \cup U_t^B) \ge \theta P(L_t^A \cup U_t^A) + (1+\theta)T_M$}
                \State Run \textbf{Procedure A}
            \Else
                \State Run \textbf{Procedure B}
            \EndIf
        \EndIf
    \EndIf
\EndWhile
\end{algorithmic}
\end{algorithm}
\begin{algorithm}[H]
\floatname{algorithm}{Procedure} 
\renewcommand{\thealgorithm}{} 
\caption{A} 
\label{proc:A}
\begin{algorithmic}[1]
\If{$T_t + P(L_t^A) + p_{s(t)}^A \le p_{l(t)}^A$}
    \If{$(1+\theta)T_t + \frac{\theta^2}{1+\theta}\big(P(L_t^A) + p_{s(t)}^A\big) \le \theta\, p_{l(t)}^A$}
        \State Schedule $J_{s(t)}^A$
    \Else
        \State Schedule $J_{l(t)}^A$
    \EndIf
\Else
    \If{$\left(1 - \frac{\theta^2}{1+\theta}\right)T_t + \frac{\theta^2}{1+\theta}p_{l(t)}^A \le \theta\big(P(L_t^A) + p_{s(t)}^A\big)$}
        \State Schedule $J_{s(t)}^A$
    \Else
        \State Schedule $J_{l(t)}^A$
    \EndIf
\EndIf
\end{algorithmic}
\end{algorithm}

\begin{algorithm}[H]
\floatname{algorithm}{Procedure} 
\renewcommand{\thealgorithm}{} 
\caption{B} 
\label{proc:B}
\begin{algorithmic}[1]
\If{$T_t + P(L_t^B) + p_{s(t)}^B \le p_{l(t)}^B$}
    \If{$(1+\theta)T_t + \frac{1}{1+\theta}\big(P(L_t^B) + p_{s(t)}^B\big) \le p_{l(t)}^B$}
        \State Schedule $J_{s(t)}^B$
    \Else
        \State Schedule $J_{l(t)}^B$
    \EndIf
\Else
    \If{$\left(\theta - \frac{1}{1+\theta}\right)T_t + \frac{1}{1+\theta}p_{l(t)}^B \le P(L_t^B) + p_{s(t)}^B$}
        \State Schedule $J_{s(t)}^B$
    \Else
        \State Schedule $J_{l(t)}^B$
    \EndIf
\EndIf
\end{algorithmic}
\end{algorithm}

\section{Competitive Analysis}
In this section, we analyze the performance guarantee of Algorithm~AOSA under the condition $1\leq \theta\leq \frac{\sqrt{5}+1}{2}$, which is equivalent to
\begin{align}\label{c1}
  \theta^2\leq \theta+1
\end{align}

For the competitive analysis, let $\bm{\sigma}$ denote the schedule generated by the online algorithm, with objective value  $Z(\sigma) = C_{\max}^A(\sigma) + \theta C_{\max}^B(\sigma)$. 
We compare $Z(\sigma)$ with the value of an optimal \textbf{offline} schedule, denoted by $\bm{\pi}$, which has objective value $Z(\pi) = C_{\max}^A(\pi) + \theta C_{\max}^B(\pi)$.

Our analysis focuses on the structural properties of $\bm{\sigma}$. 
The schedule $\bm{\sigma}$ can be partitioned into one or more \textbf{blocks}:
\begin{itemize}
    \item A \textbf{block} is a maximal set of jobs processed contiguously with no intervening idle time.
    \item A block is \textbf{pure} if it contains jobs from a single agent and \textbf{mixed} otherwise.
\end{itemize}

The following notations are introduced to analyze the structure of the last mixed block of $\bm{\sigma}$, denoted by $\bm{\mathcal{M}}$.
\begin{itemize}
    \item $\bm{\mathcal{A}}$ (resp. $\bm{\mathcal{B}}$): the final contiguous sequence of A-jobs (resp. B-jobs) in $\sigma$. 
          Note that $\mathcal{A}$ may lie outside $\mathcal{M}$ and thus form a pure block.
    \item $\bm{\mathcal{A'}}$ (resp. $\bm{\mathcal{B'}}$): the set of A-jobs (resp. B-jobs) in $\mathcal{M}$ that are scheduled before all jobs in $\mathcal{A} \cup \mathcal{B}$. 
          Either $\mathcal{A'}$ or $\mathcal{B'}$ may be empty.
    \item $\bm{J_u^A}$ (resp. $\bm{J_v^B}$): the A-job (resp. B-job) in $\mathcal{M}$ scheduled immediately before $\mathcal{B}$ (resp. $\mathcal{A}$).
    \item $\bm{\overline{\mathcal{A'}}} = \mathcal{A'} \setminus \{J_u^A\}$ if $J_u^A \in \mathcal{A'}$ and $\bm{\overline{\mathcal{B'}}} = \mathcal{B'} \setminus \{J_v^B\}$ if $J_v^B \in \mathcal{B'}$.
\end{itemize}

\begin{figure}[htbp]
\centering
\includegraphics[width=0.5\textwidth]{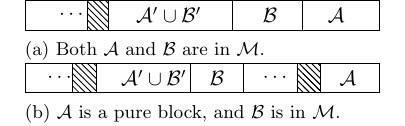}
\caption{An example of notations defined for $\sigma$.}
\label{Fig2_sec2}
\end{figure}

Figure~\ref{Fig2_sec2} illustrates the structure of the last mixed block $\mathcal{M}$ and the associated notations. 

We will now prove that the competitive ratio of Algorithm {\bf{AOSA}} is $1 + \frac{\theta}{1+\theta}$. The proof is by contradiction. Assume, to the contrary, that there exists an instance for which the competitive ratio of Algorithm {\bf{AOSA}} exceeds this value.
Let $\mathcal{I}$ be a minimal counterexample instance---that is, an instance with the minimum number of jobs that violates the bound. Let $\sigma$ be the schedule produced by {\bf{AOSA}} and $\pi$ be an optimal schedule for instance $\mathcal{I}$. Thus, we have:
\begin{align}\label{neq2}
Z(\sigma) - Z(\pi) &> \frac{\theta}{1+\theta}\,Z(\pi).
\end{align}

In the remainder of this section, we analyze the smallest counterexample by deriving a collection of structural properties of the schedule $\sigma$. 
We then employ these properties to show that $\mathcal{I}$ cannot constitute a valid counterexample. We begin by presenting a series of fundamental observations about $\sigma$.

\begin{observation} \label{observation1}
Let $T_0$ be the start time of a block in the schedule $\sigma$ produced by Algorithm AOSA. Then the following properties hold:
\begin{enumerate}[label=(\roman*)]
    \item All jobs arriving before $T_0$ in the block belong to the same agent.
    \item For each A‐job $J_{j}^{A}$ with release time $r_{j}^{A}<T_{0}$,  we have $T_0 \le \frac{\theta}{1+\theta} p_j^A$.
    \item For each B‐job $J_{j}^{B}$ with release time $r_{j}^{B}<T_{0}$,  we have $T_0 \le \frac{1}{1+\theta} p_j^B$.
\end{enumerate}
\end{observation}

\begin{observation}\label{obs2}
If the final contiguous A-job block $\mathcal{A}$ in the schedule $\sigma$ begins immediately after an idle interval, then
\begin{align}\label{neq_obs2}
C_{\max}^A(\sigma)\;\le\;\Bigl(1 + \tfrac{\theta}{1+\theta}\Bigr)\,C_{\max}^A(\pi).
\end{align}
\end{observation}

\begin{observation}\label{obs3}
If the final contiguous B-job block $\mathcal{B}$ in the schedule $\sigma$ begins immediately after an idle interval, then
\begin{align}\label{neq_obs3}
C_{\max}^B(\sigma)\;\le\;\Bigl(1 + \tfrac{\theta}{1+\theta}\Bigr)\,C_{\max}^B(\pi).
\end{align}
\end{observation}

\begin{lemma}
Both $\mathcal{A}$ and $\mathcal{B}$ belong to the mixed block $\mathcal{M}$ in the schedule $\sigma$.
\end{lemma}

\begin{proof}
We prove the lemma by contradiction. Assume, without loss of generality, that $\mathcal{A}$ does not belong to $\mathcal{M}$. This implies that $\mathcal{A}$ must constitute a pure A-block that starts after an idle time. By Observation~\ref{obs2}, this leads to inequality (\ref{neq_obs2}).

Under this assumption, the set $\mathcal{B}$ cannot start after an idle interval. If it did, Observation~\ref{obs3} would yield inequality (\ref{neq_obs3}), and combining this with (\ref{neq_obs2}) would contradict our initial assumption (\ref{neq2}). Therefore, $\mathcal{B}$ must be preceded by an $A$-job, denoted by $J_u^A$. Note that $J_u^A \in \mathcal{M}$ and $J_u^A \notin \mathcal{A}$. This structural arrangement is illustrated in the Gantt chart in Fig.~\ref{Fig1_of_Lemma1}.

\begin{figure}[htbp]
\centering
\includegraphics[width=0.5\textwidth]{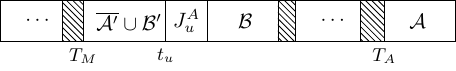}
\caption{$\mathcal{A}$ is a pure lock in $\sigma$.}
\label{Fig1_of_Lemma1}
\end{figure}

From this schedule structure, we can establish the following  makespans in $\sigma$:
\begin{align}
\label{neq-lemma1-1}
C_{\max}^A(\sigma) &= T_A + P(\mathcal{A}), \\
\label{neq-lemma1-2}
C_{\max}^B(\sigma) &= T_M + P(\mathcal{A'} \cup \mathcal{B'} \cup \mathcal{B}). 
\end{align}
Furthermore, any feasible schedule, including the optimal offline schedule $\pi$, must process these jobs. Thus, we have the trivial lower bounds:
\begin{align}
\label{neq-lemma1-3}
C_{\max}^A(\pi) &\geq P(\mathcal{A'} \cup \mathcal{A}), \\
\label{neq-lemma1-4}
C_{\max}^B(\pi) &\geq P(\mathcal{B'} \cup \mathcal{B}). 
\end{align}

    We now analyze the schedule structure by considering two cases based on the minimal release time of jobs in $\mathcal{B}$.
\begin{description}
 \item[\bf{Case 1.}] $r_{\min}(\mathcal{B}) \le t_u$.

At decision time $t_u$ (the start of job $J_u^A$) there are available jobs from both agents. 
We distinguish two subcases according to the ordering of
$r_{\min}(\overline{\mathcal{A}'}\cup U_{t_u}^A)$ and $r_{\min}(\mathcal{B}'\cup U_{t_u}^B)$.
\begin{itemize}
 \item $r_{\min}(\overline{\mathcal{A}'}\cup U_{t_u}^A)\le r_{\min}(\mathcal{B}'\cup U_{t_u}^B)$. 
 
 By Observation~\ref{observation1} every $B$-job in $\mathcal{M}$ arrives no earlier than $T_M$, hence
  \begin{align}
    \label{neq-Lemma-1-case1-1}
    C_{\max}^B(\pi)\ge T_{M}+P(\mathcal{B}'\cup\mathcal{B}).
  \end{align}
  
 Since Algorithm AOSA chooses an A-job to be processed at the decision point $t_u$, we obtain
\begin{align}\label{neq-Lemma-1-case1-2}
T_{M}+P(\mathcal{B'}\cup\mathcal{B})>\theta\bigl(P(\mathcal{A'})-T_M\bigr).
\end{align}

Combining (\ref{neq-Lemma-1-case1-1}) and (\ref{neq-Lemma-1-case1-2}),  regardless of whether 
$T_M\leq \tfrac{\theta}{1+\theta}P(\mathcal{A'})$ holds or not, we deduce that
\begin{align}
\label{neq-Lemma-1-case1-3}
C_{\max}^B(\pi)\ge \tfrac{\theta}{1+\theta}P(\mathcal{A'}).
\end{align}

Together with (\ref{neq-lemma1-3}), this yields
\begin{align}\label{neq-Lemma-1-case1-4}
Z(\pi)\ge \theta P(\mathcal{A'})+P(\mathcal{A}).
\end{align}

If $r_{\min}(\mathcal{A})<T_\mathcal{A}$, then by Observation~\ref{observation1} we have 
$T_\mathcal{A}\le \tfrac{\theta}{1+\theta}P(\mathcal{A})$. 
Together with (\ref{neq-lemma1-1}), (\ref{neq-lemma1-2}), (\ref{neq-lemma1-3}), and (\ref{neq-Lemma-1-case1-1}), this implies
\begin{align*}
Z(\sigma)-Z(\pi)
&\le T_\mathcal{A}+(\theta-1)P(\mathcal{A'}) \\
&\le \tfrac{\theta}{1+\theta}P(\mathcal{A})+\tfrac{\theta^2}{1+\theta}P(\mathcal{A'}).
\end{align*}
In view of (\ref{neq-Lemma-1-case1-4}), this contradicts (\ref{neq2}).

If $r_{\min}(\mathcal{A})\ge T_\mathcal{A}$, then
\begin{align}\label{neq-Lemma-1-case1-AB1}
C_{\max}^A(\pi)\ge T_\mathcal{A}+P(\mathcal{A}).
\end{align}
Combining this with (\ref{neq-lemma1-1}), (\ref{neq-lemma1-2}), and (\ref{neq-Lemma-1-case1-1}), we obtain
\begin{align}\label{neq-Lemma-1-case1-AB2}
Z(\sigma)-Z(\pi)\le \theta P(\mathcal{A'}).
\end{align}

Moreover, since $T_\mathcal{A}\ge T_M+P(\mathcal{A'})+P(\mathcal{B'}\cup\mathcal{B})$, it follows from 
(\ref{neq-Lemma-1-case1-1}), (\ref{neq-Lemma-1-case1-2}), and (\ref{neq-Lemma-1-case1-AB1}) that
\begin{align*}
Z(\pi)\ge (1+\theta)P(\mathcal{A'}).
\end{align*}
Together with (\ref{neq-Lemma-1-case1-AB2}), this again contradicts (\ref{neq2}).

\item $r_{\min}(\overline{\mathcal{A'}}\cup U_{t_u}^A)>r_{\min}(\mathcal{B'}\cup U_{t_u}^B)$.

In this case, by Observation~\ref{observation1}, all $A$-jobs in $\mathcal{M}$ must have release times no earlier than $T_M$. 
This implies
\begin{align}
\label{neq-Lemma-1-case1-BA1}
C_{\max}^A(\pi)\ge T_M+P(\mathcal{A'}\cup\mathcal{A}).
\end{align}

Moreover, since Algorithm AOSA schedules job $J_u^A$ at time $t_u$, we further obtain
\begin{align}\label{neq-Lemma-1-case1-BA2}
P(\mathcal{B'}\cup\mathcal{B})-T_M>\theta\bigl(T_M+P(\mathcal{A'})\bigr).
\end{align}
Together with (\ref{neq-lemma1-1}), (\ref{neq-lemma1-2}), (\ref{neq-lemma1-4}), 
(\ref{neq-Lemma-1-case1-BA1}), this yields
\begin{align}\nonumber
Z(\sigma)-Z(\pi) 
&\le T_\mathcal{A}+(\theta-1)\bigl(T_M+P(\mathcal{A'})\bigr) \\\label{delta-neq-Lemma-1-case1-BA2}
&\le T_\mathcal{A}+\tfrac{\theta-1}{\theta}P(\mathcal{B'}\cup\mathcal{B}).
\end{align}

If $r_{\min}(\mathcal{A})<T_\mathcal{A}$, then by Observation~\ref{observation1} we have 
$T_\mathcal{A}\le \tfrac{\theta}{1+\theta}P(\mathcal{A})$. 
Combining this with (\ref{delta-neq-Lemma-1-case1-BA2}), we obtain that
\begin{align*}
Z(\sigma)-Z(\pi) \le \tfrac{\theta}{1+\theta}\bigl(P(\mathcal{A})+P(\mathcal{B'}\cup\mathcal{B})\bigr).
\end{align*}
In view of (\ref{neq-lemma1-3}) and (\ref{neq-lemma1-4}), this contradicts (\ref{neq2}).

If $r_{\min}(\mathcal{A})\ge T_\mathcal{A}$, then by combining 
(\ref{neq-lemma1-1}), (\ref{neq-lemma1-2}), (\ref{neq-lemma1-4}), and (\ref{neq-Lemma-1-case1-AB1}), we obtain
\begin{align}\label{neq-Lemma-1-case1-BA3}
Z(\sigma)-Z(\pi)\le \theta\bigl(T_M+P(\mathcal{A'})\bigr).
\end{align}

Moreover, since 
$C_{\max}^A(\pi)\ge T_\mathcal{A}\ge T_M+P(\mathcal{A'})+P(\mathcal{B'}\cup\mathcal{B})$, 
it follows from (\ref{neq-Lemma-1-case1-BA2}) that
$Z(\pi)\ge (1+\theta)\bigl(T_M+P(\mathcal{A'})\bigr).$
Together with (\ref{neq-Lemma-1-case1-BA3}), this again contradicts (\ref{neq2}).
\end{itemize}

\item[\bf{Case 2.}] 
$r_{\min}(\mathcal{B}) > t_u$. In this case, all jobs in $\mathcal{B}$ arrive strictly after $t_u$, and thus $U_{t_u}^B=\emptyset$. Consequently, 
\begin{align}\label{neq-lemma1-case2-1}
C_{\max}^B(\pi) &\ge t_u+P(\mathcal{B}).
\end{align}

We further show that
\begin{align}
\label{neq-lemma1-case2-2}
t_u \;\geq\; \frac{\theta}{1+\theta}p_u^A.
\end{align}
Indeed, $r_{\min}(\mathcal{B}) > t_u$ implies that $U_{t_u}^B=\emptyset$. If inequality \eqref{neq-lemma1-case2-2} were violated, then by Algorithm~AOSA (which schedules $J_u^A$ at $t_u$), there would be at least one other available $A$-job at that time, with $J_u^A$ being the largest. In that case, all available $A$-jobs other than $J_u^A$ belong to $\mathcal{A}$, and $J_u^A$ and the remaining available $A$-jobs would be scheduled within the same block, contradicting the assumption that $\mathcal{A}$ is a pure block.

Again, we consider two subcases based on the value of $r_{\min}(\mathcal{A})$.
\begin{description}  
  \item[$\bullet$] $r_{\min}(\mathcal{A})<T_\mathcal{A}$, then by Observation~\ref{observation1} we have 
$T_\mathcal{A}\le \frac{\theta}{1+\theta}P(\mathcal{A})$. 
Combining this with (\ref{neq-lemma1-1}), (\ref{neq-lemma1-2}), (\ref{neq-lemma1-3}), and (\ref{neq-lemma1-case2-1}), we deduce
\begin{align}\nonumber
Z(\sigma)-Z(\pi) &\le T_\mathcal{A}-P(\mathcal{A}')+\theta p_u^A\\\label{neq-lemma1-case2-3}
&\le \frac{\theta}{1+\theta}\bigl(P(\mathcal{A})+\theta p_u^A\bigr).
\end{align}
The last inequality follows from $P(\mathcal{A}')\ge p_u^A$ and $\theta-1 \le \frac{\theta^2}{1+\theta}$. 

Furthermore, noting that $P(\mathcal{A}')\ge p_u^A$ and $C_{\max}^A(\pi)\ge P(\mathcal{A}'\cup \mathcal{A})$, 
and combining (\ref{neq-lemma1-case2-1}) with (\ref{neq-lemma1-case2-2}), we obtain
\begin{align*}
Z(\pi) \ge P(\mathcal{A})+\theta p_u^A.
\end{align*}
Together with (\ref{neq-lemma1-case2-3}), this leads to a contradiction with (\ref{neq2}).

  \item[$\bullet$] $r_{min}(\mathcal{A}) \geq T_\mathcal{A}$. 
  Combining (\ref{neq-lemma1-1}), (\ref{neq-lemma1-2}), (\ref{neq-Lemma-1-case1-AB1}), and (\ref{neq-lemma1-case2-1}), we obtain
\begin{align*}
Z(\sigma)-Z(\pi)\le \theta p_u^A.
\end{align*}

Furthermore, noting that $T_\mathcal{A}\ge t_u+p_u^A$, and combining (\ref{neq-Lemma-1-case1-AB1}), (\ref{neq-lemma1-case2-1}), and (\ref{neq-lemma1-case2-2}), we deduce
\begin{align*}
Z(\pi) \ge T_\mathcal{A} + \theta t_u \ge (1+\theta)p_u^A.
\end{align*}
These inequalities again contradict (\ref{neq2}).
\end{description} 
\end{description}
 
Therefore, $\mathcal{A}$ must be contained in the block $\mathcal{M}$. 
By a similar argument, $\mathcal{B}$ also belongs to the mixed block in $\sigma$. 

This completes the proof of the lemma. \hfill$\Box$
\end{proof}

By Lemma~1, both $\mathcal{A}$ and $\mathcal{B}$ must belong to the same mixed block $\mathcal{M}$.
Hence, the jobs in $\mathcal{A}\cup\mathcal{B}$ are processed consecutively in $\sigma$.
It follows that at least one of $J_u^A$ and $J_v^B$ exists.
In fact, we shall prove in the subsequent lemmas that if $\mathcal{I}$ is a counterexample, then both $J_u^A$ and $J_v^B$ must exist.

\begin{figure}[htbp]
\centering
\includegraphics[width=0.4\textwidth]{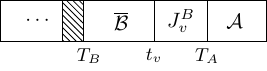}
\caption{$\mathcal{B}$ is scheduled immediately after an idle interval in $\sigma$.}
\label{Fig-sigma1}
\end{figure}

Suppose next that $\mathcal{B}$ is scheduled immediately after an idle interval in $\sigma$.
In this case, $J_v^B$ necessarily exists.
The corresponding Gantt chart is illustrated in Figure~\ref{Fig-sigma1}.
With respect to $\sigma$, we then obtain
\begin{align}\label{neq1-sigma1}
C_{\max}^A(\sigma) &= T_B + P(\mathcal{B}) + P(\mathcal{A}),\\
\label{neq3-sigma1}
C_{\max}^B(\sigma) &= T_B + P(\mathcal{B}).
\end{align}

\begin{lemma}\label{lemma2}
    If $\mathcal{B}$ starts immediately after an idle interval in $\sigma$, then $C_{max}^A(\pi)<C_{max}^B(\pi)$ must hold.    
\end{lemma}  
\begin{proof}
 Suppose, for the sake of contradiction, that in $\sigma$   $\mathcal{B}$ starts immediately after an idle interval and $C_{max}^A(\pi)\geq C_{max}^B(\pi)$. 
 We will derive a contradiction to~(\ref{neq2}).
  
  To derive lower bounds for the optimal schedule $\pi$, let $T_{\mathrm{opt}}$ denote the earliest start time, under $\pi$, of any job in $\mathcal{M}$. Then we obtain
\begin{align}\label{neq3-lemma2}
C_{max}^A(\pi)&\geq  T_{opt}+P(\mathcal{A}\cup \mathcal{B})\\
\label{neq4-lemma2}
C_{max}^B(\pi)&\geq  T_{opt}+P(\mathcal{B})
\end{align}
  
  We distinguish two cases, depending on whether $T_{\mathrm{opt}}\le T_B$ or not.
  \begin{description}
\item[$\bullet$] {\boldmath{$T_{opt}\geq T_B$}}. 
    Substituting the condition $T_{opt} \ge T_B$ into the lower bounds from (\ref{neq3-lemma2}) and (\ref{neq4-lemma2}) and comparing them with the makespans in $\sigma$ (from (\ref{neq1-sigma1}) and (\ref{neq3-sigma1})), we can obtain that $Z(\sigma) \le Z(\pi)$. This directly contradicts  (\ref{neq2}).
  
\item[$\bullet$] { \boldmath{$T_{opt} < T_B$}.}
    This condition implies that at least one job in $\mathcal{A} \cup \mathcal{B}$ is released and starts before time $T_B$.   
    
    If the job starting before $T_B$ belongs to $\mathcal{B}$, then, by Observation~\ref{observation1},
\begin{align*}
  P(\mathcal{B}) \ge (1+\theta)\, T_B .
\end{align*}
Combining this with \eqref{neq1-sigma1}, \eqref{neq3-sigma1}, \eqref{neq3-lemma2}, and \eqref{neq4-lemma2} yields
\begin{align*}
  Z(\sigma)-Z(\pi) \le P(\mathcal{B}).
\end{align*}
    Together with \eqref{neq3-lemma2} and \eqref{neq4-lemma2}, this contradicts \eqref{neq2}.
    
    If the job starting before $T_B$ belongs to $\mathcal{A}$, by Observation~\ref{observation1}, its processing time is at least $\bigl(1+\tfrac{1}{\theta}\bigr)T_B$.
    In $\pi$, all $B$-jobs start after this $A$-job. Thus,
\begin{align*}
  C_{\max}^{B}(\pi)\;\ge\; \Bigl(1+\tfrac{1}{\theta}\Bigr)T_B + P(\mathcal{B}).
\end{align*}
    Combining this with \eqref{neq1-sigma1}, \eqref{neq3-sigma1}, and \eqref{neq3-lemma2} yields
    $Z(\sigma)\le Z(\pi)$, contradicting~\eqref{neq2}.
  \end{description}  
  This completes the proof. \qed 
\end{proof}  

    Lemma~2 shows that $C_{\max}^A(\pi) < C_{\max}^B(\pi)$. 
Consequently, we obtain the following lower bounds on the optimal schedule $\pi$:
\begin{align}\label{neq1-opt}
C_{\max}^A(\pi) &\;\geq\; T_{\mathrm{opt}} + P(\mathcal{A}), \\[4pt]
\label{neq2-opt}
C_{\max}^B(\pi) &\;\geq\; T_{\mathrm{opt}} + P(\mathcal{A}) + P(\mathcal{B}),
\end{align}
where $T_{\mathrm{opt}}$ denotes the earliest starting time of the jobs in 
$\mathcal{M}$ under $\pi$.

\begin{lemma}\label{lemma3}
    If $\mathcal{B}$ is scheduled immediately after an idle interval in $\sigma$,  then it holds that $r_{\min}(\mathcal{A}) > t_v$.
\end{lemma} 
\begin{proof}   
    We argue by contradiction. Assume, for the sake of contradiction, that $r_{\min}(\mathcal{A}) \leq t_v$.  
    We distinguish two cases depending on whether $r_{\min}(\mathcal{A}) < r_{\min}(\mathcal{B})$ holds.

\begin{description}
    \item[\bf Case 1:] $r_{\min}(\mathcal{A}) < r_{\min}(\mathcal{B})$. \\[2mm]
   At time $t_v$, both some $A$-jobs and job $J_v^B$ are available. Since Algorithm AOSA gives priority to $B$-jobs, we obtain
   \begin{align*}
       \theta P(\mathcal{A}) \;\geq\; (1+\theta)T_B + P(\mathcal{B}).
   \end{align*}
   By combining this inequality with (\ref{neq1-opt}) and (\ref{neq2-opt}), we obtain
   \begin{align*}
       Z(\pi) \;\geq\; (1+\theta)T_B + \theta P(\mathcal{B}) + P(\mathcal{A}) + P(\mathcal{B}).
   \end{align*}
   This bound, together with (\ref{neq1-sigma1}) and (\ref{neq3-sigma1}), contradicts (\ref{neq2}).
   
\item[\bf Case 2:] $r_{\min}(\mathcal{A}) \geq r_{\min}(\mathcal{B})$. \\[2mm]
   Again, at time $t_v$ both $A$-jobs and $J_v^B$ are available, and Algorithm AOSA prioritizes $B$-jobs. Hence,
   \begin{align}\label{neq1-case1-lemma3}
       \theta P(\mathcal{A}) \;>\; P(\mathcal{B}) - (1+\theta)T_B .
   \end{align}
   We claim that
   \begin{align}\label{neq3-opt}
       Z(\pi) \;\geq\; (1+\theta)T_B + (1+\theta) P(\mathcal{A}) + \theta P(\mathcal{B}).
   \end{align}     
   If $T_{\text{opt}} \ge T_B$, then (\ref{neq3-opt}) follows directly from (\ref{neq1-sigma1}) and (\ref{neq3-sigma1}).  
   Otherwise, there must exist a $B$-job starting before $T_B$ in $\pi$. By Observation~\ref{observation1}, that job has processing time at least $(1+\theta)T_B$. Since all $A$-jobs start after this $B$-job in $\pi$, we have
   \begin{align*}
       C_{\max}^A(\pi) \;\geq\; (1+\theta)T_B + P(\mathcal{A}).
   \end{align*} 
   Combining this bound with (\ref{neq2-opt}) yields (\ref{neq3-opt}).  

   Finally, using (\ref{neq1-sigma1}), (\ref{neq3-sigma1}), and (\ref{neq3-opt}), we obtain 
   \[
   Z(\sigma) - Z(\pi) \;\leq\; \theta P(\mathcal{B}).
   \]
   Together with (\ref{neq1-case1-lemma3}) and (\ref{neq3-opt}), this contradicts (\ref{neq2}).    
\end{description}
    This completes the proof.
\end{proof}

   Now we are ready to establish the existence of both $J_u^A$ and $J_v^B$.  

\begin{lemma}\label{lemma4}
In schedule $\sigma$, there exists an $A$-job that immediately precedes $\mathcal{B}$, 
and, by symmetry, there exists a $B$-job that immediately precedes $\mathcal{A}$.
\end{lemma}
\begin{proof}
We argue by contradiction. 
Assume, to the contrary, that $\mathcal{B}$ is scheduled immediately after an idle interval in $\sigma$. 
Then $\mathcal{A}$ must follow $\mathcal{B}$ consecutively in $\sigma$. 
In this case, the corresponding Gantt chart of the schedule is illustrated in Fig.~\ref{Fig-sigma1}. 

From Lemmas \ref{lemma2},\ref{lemma3}, we have
$C_{max}^A(\pi)<C_{max}^B(\pi)$ and $r_{min}(\mathcal{A})>t_v$.

    Since $r_{\min}(\mathcal{A})>t_v$, it follows that  
\begin{align}\label{neq1-opt-lemma4}
C_{\max}^A(\pi)\;\geq\; t_v+P(\mathcal{A}).
\end{align}
Combining this with \eqref{neq1-sigma1}, \eqref{neq3-sigma1}, and \eqref{neq2-opt}, we obtain  
\begin{align}\label{neq2-opt-lemma4}
Z(\sigma)-Z(\pi)\;\leq\;\theta \,(T_B-T_{\text{opt}})+p_v^B.
\end{align} 

    Since $J_v^B$ is the last $B$-job and no $A$-jobs are available at time $t_v$, it follows directly from Observation~\ref{observation1} that
\begin{align}\label{neq10-lemma2}
t_v\geq \frac{1}{1+\theta}p_v^B.
\end{align} 
    
   We now distinguish two cases according to the value of $C_{\max}^B(\pi)$.
\begin{description}
\item[\bf Case 1:] $C_{\max}^B(\pi)\geq t_v+p_v^B$.  
  Combining this with \eqref{neq1-sigma1}, \eqref{neq3-sigma1}, and \eqref{neq1-opt-lemma4}, we deduce that  
  \begin{align*}
  Z(\sigma)-Z(\pi)\;\leq\; p_v^B.
  \end{align*}
  Moreover, together with $C_{\max}^B(\pi)\geq t_v+p_v^B$, \eqref{neq1-opt-lemma4}, and \eqref{neq10-lemma2}, this leads to a contradiction with \eqref{neq2}.  
  
\item[\bf Case 2:] $C_{\max}^B(\pi)< t_v+p_v^B$.  
  In this case, job $J_v^B$ must start before $t_v$ in $\pi$.  
  We claim that  
  \begin{align}\label{neq1-claim-lemma4}
  C_{\max}^A(\pi)\;<\; T_B+p_v^B.
  \end{align}
  Otherwise, if $C_{\max}^A(\pi)\geq T_B+p_v^B$, then together with $C_{\max}^B(\pi)\geq C_{\max}^A(\pi)$ and \eqref{neq2-opt-lemma4}, we obtain a contradiction with \eqref{neq2}.
    
    Since $J_v^B$ starts before $t_v$ and $r_{\min}(\mathcal{A})>t_v$, i.e., all jobs in $\mathcal{A}$ are scheduled after $J_v^B$ and $t_v$ in $\pi$, we have  
\begin{align}\label{neq3-opt-lemma4}
C_{\max}^A(\pi)\;\geq\;\max\{p_v^B,\,t_v\}+P(\mathcal{A}). 
\end{align}    
    
    Note that $t_v = T_B + P(\mathcal{B}) - p_v^B$, 
$C_{\max}^B(\pi)\geq P(\mathcal{B})$, and 
$P(\mathcal{B}) = P(\overline{\mathcal{B}})+p_v^B$.  
Combining these with \eqref{neq1-sigma1}, \eqref{neq3-sigma1}, and \eqref{neq3-opt-lemma4}, we deduce that  
\begin{align}\label{neq12-lemma4}
Z(\sigma)-Z(\pi)&\;\leq\; \min\{t_v,\,p_v^B\}+\theta T_B,\\
Z(\pi)&\;\geq\; \theta P(\overline{\mathcal{B}})+\theta p_v^B+\max\{t_v,\,p_v^B\}.
\end{align}
    
    Moreover, we note the following two properties of $J_v^B$:  
\begin{itemize}
  \item[(i)] $J_v^B$ starts before $T_B$ in $\pi$.  
  This holds because $J_v^B$ starts before the jobs in $\mathcal{A}$ (as $r_{\min}(\mathcal{A})>t_v$), and by \eqref{neq1-claim-lemma4} its completion time cannot exceed $T_B+p_v^B$.  
  \item[(ii)] $J_v^B$ is the largest $B$-job in $\mathcal{B}$.  
  This follows since \eqref{neq1-claim-lemma4} and \eqref{neq3-opt-lemma4}, together with $t_v = T_B+P(\mathcal{B})-p_v^B$, imply that $p_v^B$ dominates the processing times of all other $B$-jobs.  
\end{itemize}
     
    Since no $A$-jobs are available while the jobs in $\mathcal{B}$ are being scheduled, 
and Procedure~B does not assign priority to the largest $B$-job $J_v^B$, 
we distinguish two subcases.
\begin{description}
 \item[\bf Case 2.1:] $T_B+P(\overline{\mathcal{B}})\leq p_v^B$.  
  By Procedure~B, when the job immediately preceding $J_v^B$ is scheduled, it must hold that
  \begin{align*}
     (1+\theta)T_B+\frac{1}{1+\theta}P(\overline{\mathcal{B}})\;\leq\; p_v^B.
  \end{align*}   
  Combining this with \eqref{neq3-opt-lemma4} and the fact that
 $C_{\max}^B(\pi)\ge P(\overline{\mathcal{B}})+p_v^B$, we obtain
  \begin{align*}\nonumber
  Z(\pi)&\;\geq\;\theta P(\overline{\mathcal{B}})+\theta p_v^B+p_v^B \\ 
        &\;\geq\;(1+\theta)\big((1+\theta)T_B+P(\overline{\mathcal{B}})\big).
  \end{align*}
  Together with $t_v = T_B + P(\overline{\mathcal{B}})$ and \eqref{neq12-lemma4}, 
this contradicts \eqref{neq2}.
    
  \item[\bf{Case 2.2:}] $T_B+P(\mathcal{\overline{B}})> p_v^B$. By Procedure $B$, when the job before $J_v^B$ is scheduled, we have
\begin{align*}
   (\theta-\frac{1}{1+\theta})T_B+\frac{1}{1+\theta}p_v^B\leq P(\mathcal{\overline{B}}).
\end{align*}   
     Combined this with (\ref{neq3-opt-lemma4}) and $C_{max}^B(\pi)\geq P(\mathcal{\overline{B}})+p_v^B$, $t_v=T_B+P(\mathcal{\overline{B}})$, we have
\begin{align*}
Z(\pi)&\geq (1+\theta)(\theta T_B+p_v^B).
\end{align*}
 which along with  (\ref{neq12-lemma4}), leads a contradiction with (\ref{neq2}).
\end{description}      
\end{description}

    By symmetry, the second statement follows upon swapping $\mathcal{A}$ and $\mathcal{B}$. 
 
 This completes the proof. \qedhere
\end{proof}

\begin{figure}[htbp]
\centering
\includegraphics[width=0.5\textwidth]{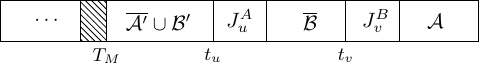}
\caption{The Gantt chart of schedule in which $J_u^A\not\in \mathcal{A}$.}
\label{Fig-sigma2}
\end{figure}

By Lemma~\ref{lemma4}, the jobs $J_u^A$ and $J_v^B$, which are processed immediately before $\mathcal{B}$ and $\mathcal{A}$ in $\sigma$, always exist.  
If $J_u^A\notin\mathcal{A}$, then $J_v^B\in\mathcal{B}$; the corresponding Gantt chart of $\sigma$ is shown in Fig.~\ref{Fig-sigma2}.  
In this case, the following equalities hold for the two agents' makespans in $\sigma$:  
\begin{align}
\label{neq1-sigma2}
C_{\max}^A(\sigma) &= T_M + P\bigl(B'\cup A'\cup B\cup A\bigr),\\
\label{neq2-sigma2}
C_{\max}^B(\sigma) &= T_M + P\bigl(B'\cup A'\cup B\bigr).
\end{align}
In what follows, we derive further conclusions under the assumption that $\mathcal{I}$ is a counterexample.

\begin{lemma}\label{lemma5}
If $J_u^A\notin\mathcal{A}$ and $r_{\min}(\mathcal{B})\leq t_u$, then $C_{\max}^A(\pi)<C_{\max}^B(\pi)$.
\end{lemma}
\begin{proof}
Assume, for contradiction, that $C_{\max}^A(\pi)\geq C_{\max}^B(\pi)$ while $J_u^A\notin\mathcal{A}$ and $r_{\min}(\mathcal{B})\leq t_u$ in $\sigma$.
We will derive a contradiction to~\eqref{neq2}.
   
   From $C_{\max}^A(\pi)\ge C_{\max}^B(\pi)$, we obtain
\begin{align}\label{neq1-opt-lemma5}
    Z(\pi)\;\ge\; P\!\bigl(\mathcal{B}'\cup\mathcal{B}\cup\mathcal{A}'\cup\mathcal{A}\bigr)\;+\;\theta\, P\!\bigl(\mathcal{B}'\cup\mathcal{B}\bigr).
\end{align}
Combining this with \eqref{neq1-sigma2} and \eqref{neq2-sigma2} yields
\begin{align}\label{neq2-opt-lemma5}
    Z(\sigma)-Z(\pi)\;\le\; (1+\theta)\,T_{M}\;+\;\theta\, P(\mathcal{A}').
\end{align}

We next distinguish two cases according to the relative order between $r_{\min}(\mathcal{A}\cup \mathcal{A}')$ and $\mathcal{B}\cup \mathcal{B}')$.
\begin{description}
   \item[\bf{Case 1:}] $r_{min}(\mathcal{A}\cup \mathcal{A}')\geq  r_{min}(\mathcal{B}\cup \mathcal{B}')$.\\
  In this case, since $J_u^A$ is scheduled at time $t_u$, the algorithm implies that
\begin{align*}
  \theta P(\mathcal{A}') \leq P(\mathcal{B}\cup \mathcal{B}')-(1+\theta)T_M.
\end{align*}
    Together with \eqref{neq1-opt-lemma5} and \eqref{neq2-opt-lemma5}, this contradicts \eqref{neq2}.
  
  \item[\bf{Case 2:}] $r_{min}(\mathcal{A}\cup \mathcal{A}')<  r_{min}(\mathcal{B}\cup \mathcal{B}')$. \\
  In this case, since $J_u^A$ is scheduled at time $t_u$, the algorithm guarantees
\begin{align}\label{neq1-case2-lemma5}
  \theta P(\mathcal{A}') \leq P(\mathcal{B}\cup \mathcal{B}')+(1+\theta)T_M.
\end{align}    

    We now claim that
\begin{align}\label{neq3-opt-lemma5}
    Z(\pi)\geq (1+\theta)T_M+P(\mathcal{B'}\cup\mathcal{B}\cup\mathcal{A'}\cup\mathcal{A})+\theta P(\mathcal{B'}\cup\mathcal{B}).
\end{align}  
    Indeed, if no jobs start before $T_M$ in $\pi$, then \eqref{neq3-opt-lemma5} holds immediately.
    Otherwise, there is a job starting before $T_M$ in $\pi$. By Observation~\ref{observation1} and the initial condition of this case,
    the job must be an $A$-job with processing time no less than $\frac{\theta+1}{\theta}T_M$.
    Since all $B$-jobs would be scheduled after this $A$-job, it follows that
\begin{align*}
C_{max}^B(\pi)\geq P(\mathcal{B'}\cup\mathcal{B})+\frac{\theta+1}{\theta}T_M.
\end{align*}
    This, together with the fact that $C_{max}^A(\pi)\geq P(\mathcal{B'}\cup\mathcal{B}\cup\mathcal{A'}\cup\mathcal{A})$,
    implies that (\ref{neq3-opt-lemma5}) holds.
   
   Finally, combining \eqref{neq3-opt-lemma5} with \eqref{neq1-sigma2} and \eqref{neq2-sigma2} yields
\begin{align*}
  Z(\sigma)-Z(\pi) \le \theta\, P(\mathcal{A}').
\end{align*}
Together with \eqref{neq1-case2-lemma5} and \eqref{neq3-opt-lemma5}, this contradicts \eqref{neq2}.
\end{description}

\medskip
In both cases we reach a contradiction. Hence $C_{\max}^A(\pi) < C_{\max}^B(\pi)$, as claimed.
This completes the proof. \qed
\end{proof}

Lemma~5 implies that the following lower bounds always hold for $\pi$:
\begin{align} \label{neq2-opt-lemma6}
C_{max}^A(\pi) &\geq P(\mathcal{A}'\cup \mathcal{A}), \\ \label{neq3-opt-lemma6}
C_{max}^B(\pi) &\geq P(\mathcal{A}'\cup \mathcal{A}) + P(\mathcal{B}'\cup \mathcal{B}).
\end{align}

\begin{lemma}\label{lemma6}
If $J_u^A\notin \mathcal{A}$ and $r_{\min}(\mathcal{B})\le t_u$, then $r_{\min}(\mathcal{A})> t_v$.
\end{lemma}
\begin{proof}
Suppose, for contradiction, that $r_{\min}(\mathcal{A})\le t_v$.
By Observation~\ref{observation1}, at most one agent has a job that starts before $T_M$ in $\pi$.
We distinguish two cases according to whether a job in $\mathcal{M}$ starts before $T_M$ in $\pi$.

\begin{description}
\item[\bf Case 1:] There exists an $A$-job that starts before $T_M$ in $\pi$. 
By Observation~\ref{observation1}, it follows that 
$r_{\min}(\mathcal{A}'\cup\mathcal{A}) < r_{\min}(\mathcal{B}'\cup\mathcal{B})$. 
Since at time $t_v$ job $J_v^B$ is scheduled first, by the rule of Algorithm~AOSA we obtain
\begin{align}\label{neq02-lemma6}
  P(\mathcal{B}'\cup\mathcal{B}) 
  \le \theta P(\mathcal{A}'\cup\mathcal{A}) - (1+\theta)T_M.
\end{align}

Combining \eqref{neq02-lemma6} with \eqref{neq1-sigma2}, \eqref{neq2-sigma2}, \eqref{neq2-opt-lemma6} and \eqref{neq3-opt-lemma6},
we get
\begin{align*}
  Z(\sigma)-Z(\pi)\le \theta P(\mathcal{A}').
\end{align*}
This, together with \eqref{neq2-opt-lemma6} and \eqref{neq3-opt-lemma6}, contradicts \eqref{neq2}.

\item[\bf Case 2:] All $A$-jobs start at or after $T_M$ in $\pi$.
We claim that
\begin{align}\label{neq1-optLB-lemma6}
  Z(\pi)\ge \theta P(\mathcal{B}'\cup\mathcal{B})+(1+\theta)P(\mathcal{A}'\cup\mathcal{A})+(1+\theta)T_M.
\end{align}
Indeed, if all $B$-jobs also start at or after $T_M$ in $\pi$, then \eqref{neq1-optLB-lemma6} follows immediately from \eqref{neq1-sigma2} and \eqref{neq2-sigma2}.
Otherwise, there exists a $B$-job that starts before $T_M$ in $\pi$. By Observation~\ref{observation1}, the processing time of this $B$-job is at least $(1+\theta)T_M$.
Since all $A$-jobs start after this $B$-job in $\pi$, we obtain
\begin{align*}
  C_{\max}^A(\pi)\ge (1+\theta)T_M + P(\mathcal{A}'\cup\mathcal{A}),
\end{align*}
which, together with \eqref{neq3-opt-lemma6} implies that \eqref{neq1-optLB-lemma6} holds.

Combining \eqref{neq1-optLB-lemma6} with \eqref{neq1-sigma2} and \eqref{neq2-sigma2}, we obtain
\begin{align}\label{delta2-lemma6}
  Z(\sigma)-Z(\pi)\le \theta P(\mathcal{B}'\cup\mathcal{B}).
\end{align}

Note that at time $t_v$, job $J_v^B$ is scheduled prior to the available jobs in $\mathcal{A}$.
According to the scheduling rule of Algorithm~AOSA, this implies that
\begin{align*}
  P(\mathcal{B}'\cup\mathcal{B}) \le (1+\theta)T_M + \theta P(\mathcal{A}'\cup\mathcal{A}).
\end{align*}
Combining this with \eqref{neq1-optLB-lemma6}, we obtain that $Z(\pi)\ge (1+\theta)\,P(\mathcal{B}'\cup\mathcal{B})$.
This, combined with \eqref{delta2-lemma6}, contradicts \eqref{neq2}.
\end{description}

Hence the assumption is false and $r_{\min}(\mathcal{A})>t_v$. This completes the proof.
\qed
\end{proof}

\begin{lemma}\label{lemma7}
If $r_{\min}(\mathcal{B})\le t_u$, then $J_u^A\in \mathcal{A}$.
\end{lemma}

\begin{proof}
We argue by contradiction.
Suppose, to the contrary, that $J_u^A\notin \mathcal{A}$ while $r_{\min}(\mathcal{B})\le t_u$.
The condition $J_u^A\notin \mathcal{A}$ implies that, in $\sigma$, all jobs in $\mathcal{A}$ are processed after $\mathcal{B}$.
In this case, the Gantt chart of $\sigma$ is shown in Fig.~\ref{Fig-sigma2}.

By Lemmas~\ref{lemma5} and~\ref{lemma6}, we have $C_{\max}^A(\pi)<C_{\max}^B(\pi)$ and $r_{\min}(\mathcal{A})>t_v$.
By the fact that $t_v = T_M + P(\mathcal{A}') + P(\mathcal{B}'\cup \mathcal{B}) - p_v^B$ and $r_{\min}(\mathcal{A})>t_v$, we obtain
\begin{align}\label{neq2-optLB-lemma7}
C_{\max}^A(\pi)
   &\ge T_M + P(\mathcal{A}'\cup \mathcal{A}) + P(\mathcal{B}'\cup \mathcal{B}) - p_v^B .
\end{align}
Combining this with \eqref{neq1-sigma2}, \eqref{neq2-sigma2}, and \eqref{neq3-opt-lemma6}, we deduce that
\begin{align}\label{neq1-delta-lemma7}
  Z(\sigma)-Z(\pi) \le p_v^B + \theta\, T_M.
\end{align}

Note that at time $t_v$, the only available job is $J_v^B$.
By Observation~\ref{observation1}, we have
\begin{align}\label{neq3-case2-lemma7}
  t_v \ge \frac{1}{1+\theta}\, p_v^B.
\end{align}

In what follows, we distinguish three cases according to the start time of $J_v^B$ in $\pi$.   
 \begin{description}
\item[\bf Case 1:] $J_v^B$ starts at or after $t_v$ in $\pi$.

In this case, $C_{\max}^B(\pi)\ge C_{\max}^B(\sigma)$.
Combining this with \eqref{neq1-sigma2}, \eqref{neq2-optLB-lemma7}, we obtain
\begin{align}\label{neq1-case1-lemma7}
  Z(\sigma)-Z(\pi) \le p_v^B.
\end{align}
Moreover, it follows from $C_{\max}^A(\pi)>t_v$ and $C_{\max}^B(\pi)\ge t_v+p_v^B$, together with \eqref{neq3-case2-lemma7}, that
\[
  Z(\pi)\ge (1+\theta)\,p_v^B.
\]
Together with \eqref{neq1-case1-lemma7}, this contradicts \eqref{neq2}.

  \item[\bf Case 2:] $J_v^B$ starts in $\pi$ between $T_M$ and $t_v$.  
  
Since $r_{\min}(\mathcal{A})>t_v$, we obtain
\[
  C_{\max}^A(\pi)\;\ge\; T_M + p_v^B.
\]
Combining this with the inequality $C_{\max}^A(\pi)<C_{\max}^B(\pi)$, it follows that
\begin{align}
  Z(\pi) \;\ge\; (1+\theta)\big(T_M+p_v^B\big).
\end{align}
This, together with \eqref{neq1-delta-lemma7},  contradict \eqref{neq2}.

\item[\bf{Case 3:}] $J_v^B$ starts before $t_M$ in $\pi$. 

This implies that $r_v^B<T_M$ and  all $A$-jobs start after $J_v^B$ in $\pi$. Therefore,
\begin{align}\label{neq1-case23-lemma7}
   C_{\max}^A(\pi) \;\ge\; p_v^B + P(\mathcal{A}'\cup \mathcal{A}).
\end{align}
 By Observation~\ref{observation1},  $r_v^B<T_M$ implies that
\begin{align}\label{neq0-case23-lemma7}
   p_v^B \;\ge\; (1+\theta)T_M.
\end{align}

Define $x := P(\mathcal{B}'\cup \mathcal{B}) - p_v^B$. Then
\[
   C_{\max}^B(\pi)\;\ge\; p_v^B+x,
\]
which, together with \eqref{neq1-case23-lemma7}, implies
\begin{align}\label{neq3-optLB-lemma7}
   Z(\pi) \;\ge\; (1+\theta)p_v^B+\theta x.
\end{align}

Moreover, combining \eqref{neq1-sigma2}, \eqref{neq2-sigma2}, \eqref{neq3-opt-lemma6}, and \eqref{neq1-case23-lemma7}, we obtain
\begin{align}\label{neq2-delta-lemma7-pre}
   Z(\sigma)-Z(\pi)\;\le\; (1+\theta)T_M+x.
\end{align}

If $x \ge p_v^B$, then combining \eqref{neq1-delta-lemma7}, \eqref{neq0-case23-lemma7}, and \eqref{neq3-optLB-lemma7},  yields a contradiction to \eqref{neq2}.

If $x < p_v^B$, then $J_v^B$ is the largest $B$-job.  
Since no $A$-job is available when the $B$-job immediately preceding $J_v^B$ is scheduled, Procedure~B ensures
\begin{align*}
   (1+\theta)T_M+\frac{1}{1+\theta}x \;\le\; p_v^B.
\end{align*}
Combining this with \eqref{neq2-delta-lemma7-pre}, we derive
\begin{align}\label{neq5-case23-lemma7}
   Z(\sigma)-Z(\pi) \;\le\; p_v^B+\frac{\theta}{1+\theta}x,
\end{align}
which, together with \eqref{neq3-optLB-lemma7}, again yields a contradiction with \eqref{neq2}.

\end{description}    
This completes the proof. \qed
\end{proof}

By an analogous argument, we obtain the following lemma.

\begin{lemma}\label{lemma8}
In schedule $\sigma$, if $r_{\min}(\mathcal{A})\le t_v$, then $J_v^B\in\mathcal{B}$.
\end{lemma}

\begin{figure}[htbp]
\centering
\includegraphics[width=0.5\textwidth]{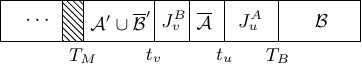}
\caption{The Scenario  with $r_{min}(\mathcal{B})>t_u$ and $J_u^A\in A$ in $\sigma$.}
\label{Fig_sigma3}
\end{figure}

\begin{lemma}
If $r_{\min}(\mathcal{B})> t_u$, then $J_u^A\notin \mathcal{A}$.
\end{lemma}
\begin{proof}
We argue by contradiction.
Suppose that $r_{\min}(\mathcal{B})>t_u$ and $J_u^A\in\mathcal{A}$.
Under this assumption, $J_v^B\notin\mathcal{B}$; hence all jobs in $\mathcal{B}$ appear at the tail of the schedule $\sigma$.
In this case, the Gantt chart of $\sigma$ is shown in Fig.~\ref{Fig_sigma3}. Thus,
\begin{align}\label{neq1-sigma3}
  Z(\sigma)=t_v+p_v^B+P(\mathcal{A})+\theta\bigl(t_v+p_v^B+P(\mathcal{A})+P(\mathcal{B})\bigr).
\end{align}

By Lemma~\ref{lemma8}, the condition $J_v^B\notin\mathcal{B}$ implies $r_{\min}(\mathcal{A})>t_v$.
Together with $r_{\min}(\mathcal{B})>t_u$, we obtain
\begin{align}
   C_{\max}^A(\pi) \;\ge\; t_v + P(\mathcal{A}) \label{neq1-opt-lemma9}\\[6pt]
   C_{\max}^B(\pi) \;\ge\; t_u + P(\mathcal{B}). \label{neq2-opt-lemma9}
\end{align}
Combining \eqref{neq1-sigma3}, \eqref{neq1-opt-lemma9}, and \eqref{neq2-opt-lemma9} with the identity
$t_u = t_v + p_v^B + P(\mathcal{A})$, we deduce
\begin{align}\label{neq3-delta-lemma9}
   Z(\sigma) - Z(\pi) \;\le\; \theta\, p_u^A + p_v^B.
\end{align}

We now distinguish two cases based on the relative order of $C_{\max}^A(\pi)$ and $C_{\max}^B(\pi)$.
\begin{description}  

\item[\bf Case 1:] $C_{\max}^A(\pi)\geq C_{\max}^B(\pi)$.

If $J_u^A$ starts after all $B$-jobs in $\pi$, then, since $r_{\min}(\mathcal{B})>t_u$, we have
\begin{align}\label{neq3-opt-lemma9}
   C_{\max}^A(\pi) \;>\; t_u + p_u^A.
\end{align}
Recalling that $t_u = t_v + p_v^B + P(\mathcal{A})$ and combining
\eqref{neq3-opt-lemma9} with \eqref{neq1-sigma3} and \eqref{neq2-opt-lemma9}, we obtain
\begin{align}\label{neq2-delta-lemma9}
   Z(\sigma) - Z(\pi) \;\le\; \theta\,p_u^A.
\end{align}
At time $t_u$, $J_u^A$ is the only available job. By Observation~\ref{observation1},
\[
   t_u \;\ge\; \tfrac{\theta}{1+\theta}\,p_u^A,
\]
which, together with \eqref{neq2-opt-lemma9}, \eqref{neq3-opt-lemma9}, and \eqref{neq2-delta-lemma9}, yields a contradiction to \eqref{neq2}.

If, instead, $J_u^A$ is scheduled before some $B$-jobs in $\pi$, then
\begin{align*}
   C_{\max}^B(\pi) \;\ge\; p_u^A + p_v^B.
\end{align*}
Combining this with $C_{\max}^A(\pi)\geq C_{\max}^B(\pi)$ and \eqref{neq3-delta-lemma9},
we again obtain a contradiction to \eqref{neq2}.

\item[\bf{Case 2:}] $C_{\max}^A(\pi) < C_{\max}^B(\pi)$. 

If $J_v^B$ starts at or after $t_v$ in $\pi$, then, since $r_{\min}(\mathcal{A}\cup\mathcal{B}) > t_v$, we have
\begin{align}\label{neq4-opt-lemma9}
   C_{\max}^B(\pi) \;\ge\; t_v + p_v^B + P(\mathcal{A}) + P(\mathcal{B}).
\end{align}
This together with \eqref{neq1-sigma3} and \eqref{neq1-opt-lemma9}, implies that
\begin{align}\label{neq4-delta-lemma9}
   Z(\sigma) - Z(\pi) \;\le\; p_v^B.
\end{align}
At time $t_v$, $J_v^B$ is the only available job. Hence, by Observation~\ref{observation1}, 
\[
   t_v \;\ge\; \tfrac{1}{1+\theta}p_v^B,
\]
which, together with \eqref{neq1-opt-lemma9}, \eqref{neq4-opt-lemma9}, and \eqref{neq4-delta-lemma9}, contradicts \eqref{neq2}.

If instead $J_v^B$ starts before $t_v$ in $\pi$, then since $r_{\min}(\mathcal{A}) > t_v$, we obtain
\[
   C_{\max}^A(\pi) \;\ge\; p_v^B + p_u^A.
\]
Combining this with $C_{\max}^A(\pi) < C_{\max}^B(\pi)$ and \eqref{neq3-delta-lemma9}, 
we again obtain a contradiction to \eqref{neq2}.

\end{description}
This completes the proof. \qed
\end{proof}

    By combining the previous three lemmas, we obtain the following results.  

\begin{lemma}\label{lemma10}
$r_{\min}(\mathcal{B}) > t_u$ holds if and only if $J_u^A \notin \mathcal{A}$.
\end{lemma}

\begin{lemma}\label{lemma11}
$r_{\min}(\mathcal{A}) > t_v$ holds if and only if $J_v^B \notin \mathcal{B}$.
\end{lemma}

By Lemmas~\ref{lemma10} and \ref{lemma11}, we have that $J_u^A \notin \mathcal{A}$ and $J_v^B \in \mathcal{B}$ whenever $r_{\min}(\mathcal{B}) > t_u$.  
In this case, the Gantt chart of schedule $\sigma$ is illustrated in Figure~\ref{Fig-sigma2}.  
Combining \eqref{neq1-sigma2} and \eqref{neq2-sigma2}, we obtain
\begin{align}\label{z-sigma3}
   Z(\sigma) \;=\; (1+\theta)T_M + P(\mathcal{A}) 
     + (1+\theta)\big(P(\mathcal{B}\cup\mathcal{B}') + P(\mathcal{A}')\big).
\end{align}
The following lemmas are established under this setting.

\begin{lemma}\label{lemma12}
If $r_{\min}(\mathcal{B}) > t_u$, then $C_{\max}^A(\pi) \;\ge\; C_{\max}^B(\pi)$.
\end{lemma}
\begin{proof}
We argue by contradiction.  
Suppose instead that $C_{\max}^A(\pi) < C_{\max}^B(\pi)$, which implies
\begin{align}\label{z1-opt-lemma12}
   Z(\pi) \;\ge\; (1+\theta) P(\mathcal{A}\cup \mathcal{A}') + \theta P(\mathcal{B}\cup \mathcal{B}').
\end{align}
By Lemmas~\ref{lemma10} and \ref{lemma11}, the condition $r_{\min}(\mathcal{B}) > t_u$ implies that 
$J_u^A \notin \mathcal{A}$, $J_v^B \in \mathcal{B}$, and $r_{\min}(\mathcal{A}) \le t_v$.

We consider two cases based on the relationship between $r_{\min}(\mathcal{A}\cup \mathcal{A}')$ and $r_{\min}(\mathcal{B}\cup \mathcal{B}')$.

\begin{description}
\item[\bf{Case 1:}] $r_{\min}(\mathcal{A}\cup \mathcal{A}') < r_{\min}(\mathcal{B}\cup \mathcal{B}')$.  

At time $t_v$, since $J_v^B$ is given priority over $A$-jobs by Algorithm~AOSA, we have
\begin{align}\label{neq1-case1-lemma12}
   P(\mathcal{B}\cup \mathcal{B}') \;\le\; \theta P(\mathcal{A}\cup \mathcal{A}') - (1+\theta) T_M.
\end{align}

Combining \eqref{z-sigma3}--\eqref{neq1-case1-lemma12}, we obtain
\[
   Z(\sigma) - Z(\pi) \;\le\; \theta P(\mathcal{A}').
\]
This, together with \eqref{z1-opt-lemma12}, contradicts \eqref{neq2}.
   
\item[\bf{Case 2:}] $r_{min}(A\cup A')\geq  r_{min}(B\cup B')$.   

     At time $t_v$, $J_v^B$ has  priority over $A$-jobs under Algorithm~AOSA. Thus, we have
\begin{align}\label{neq1-case2-lemma12}
  &P(B\cup B')\leq \theta P(A\cup A')+(1+\theta)T_M.
\end{align}

    We claim that
\begin{align}\label{z2-opt-lemma12}
    Z(\pi)\geq (1+\theta)T_M+(1+\theta)P(\mathcal{A}\cup\mathcal{A'})+\theta P(\mathcal{B}\cup\mathcal{B'}).
\end{align}

    Indeed, if a $B$-job starts before $T_M$, by Observation \ref{observation1}, the $B$-job has a processing time no less than $(1+\theta)T_M$
    and all $A$-jobs start after that $B$-job in $\pi$, which implies that
\[ C_{\max}^A(\pi) \geq (1+\theta)T_M + P(\mathcal{A}\cup \mathcal{A}').\] 

    This together with $C_{\max}^B(\pi) \geq P(\mathcal{A}\cup \mathcal{A}') + P(\mathcal{B}\cup \mathcal{B}')$, yields \eqref{z2-opt-lemma12}. 
    Otherwise, no job starts before $T_M$, and \eqref{z2-opt-lemma12} holds trivially.
    
    Combining (\ref{z-sigma3}) and (\ref{z2-opt-lemma12}), we obtain that
\[
  Z(\sigma)-Z(\pi)\leq P(B\cup B').
\]
    This, together  with \eqref{neq1-case2-lemma12} and \eqref{z2-opt-lemma12}, yields a contradiction. 
  \end{description}  
This completes the proof. \qed   
\end{proof}

By Lemma~\ref{lemma12}, with respect to $\pi$, whenever $r_{\min}(\mathcal{B}) > t_u$, we have
\begin{align}\label{ALB_opt_lemma12}
   C_{\max}^A(\pi) &\;\ge\; P(\mathcal{B}'\cup \mathcal{B}) + P(\mathcal{A}'\cup \mathcal{A}),\\
\label{BLB_opt_lemma12}
   C_{\max}^B(\pi) &\;\ge\; P(\mathcal{B}'\cup \mathcal{B}),\\
\label{BLB2_opt_lemma12}
   C_{\max}^B(\pi) &\;\ge\; t_u + P(\mathcal{B}).
\end{align}

Combining these inequalities with \eqref{neq1-sigma2} and \eqref{neq2-sigma2}, we obtain
\begin{align}\label{delta_sigam_opt_lemma12}
   Z(\sigma) - Z(\pi) \;\le\; T_M + \theta p_u^A.
\end{align}


\begin{lemma}\label{lemma13}
If $r_{\min}(\mathcal{B}) > t_u$, then $J_u^A$ is largest job among those $A$-jobs arriving at or before time $t_u$.
\end{lemma}
\begin{proof}
When $r_{\min}(\mathcal{B})>t_u$, the Gantt chart of $\sigma$ is depicted in Fig.~\ref{Fig-sigma2}.
Suppose, for contradiction, that among the $A$-jobs released no later than $t_u$ there exists an job other than $J_u^A$ with processing time at least $p_u^A$.
Denote the largest such $A$-job by $J_l^A$. Then
\begin{align}\label{lba-opt-lemma13}
   C_{\max}^A(\pi) \;\ge\; 2\,p_u^A.
\end{align}
In $\sigma$, regardless of whether $J_l^A$ is processed before or after $t_u$, we have
\begin{align}\label{neq1-lemma13}
   t_u \;\ge\; \frac{\theta}{1+\theta}\,p_u^A.
\end{align} 

We consider three cases as follows:
\begin{description}
\item[\bf Case 1:]No job starts before $T_M$ in $\pi$.
Since $C_{\max}^A(\pi)\ge C_{\max}^B(\pi)$, we have
\begin{align}\label{case1-cmaxA}
  C_{\max}^A(\pi) \;\ge\; T_M + P(\mathcal{B}'\cup \mathcal{B}) + P(\mathcal{A}'\cup \mathcal{A}).
\end{align}
Combining \eqref{case1-cmaxA} with \eqref{neq1-sigma2}, \eqref{neq2-sigma2}, and \eqref{BLB2_opt_lemma12}, we obtain
\[
  Z(\sigma) - Z(\pi) \;\le\; \theta\, p_u^A,
\]
which, together with \eqref{BLB2_opt_lemma12}, \eqref{lba-opt-lemma13}, and \eqref{neq1-lemma13}, contradicts \eqref{neq2}.

\item[\bf Case 2:] One of $J_u^A$ and $J_l^A$ starts before $T_M$. 
By Observation~\ref{observation1}, that $A$-job has processing time at least $(1+\tfrac{1}{\theta})T_M$. Hence
\[
   C_{\max}^A(\pi) \;\ge\; \frac{1+\theta}{\theta} T_M + p_u^A.
\]
Since all $B$-jobs start after this $A$-job in this case, we have $C_{\max}^B(\pi) \ge p_u^A$. Combining these bounds, we obtain
\[
   Z(\pi) \;\ge\; (1+\theta) p_u^A + \frac{1+\theta}{\theta} T_M.
\]
Together with \eqref{delta_sigam_opt_lemma12}, this leads to a contradiction.

\item[\bf Case 3:] There exists a job other than $J_u^A$ and $J_l^A$ that starts before $T_M$ in $\pi$.
By Observation~\ref{observation1}, this job has a processing time  at least $(1+\tfrac{1}{\theta})T_M$.
Consequently, both $J_u^A$ and $J_l^A$ start after that job.
Hence
\[
   C_{\max}^A(\pi) \ge\frac{1+\theta}{\theta} T_M + 2 p_u^A, 
\]
which together with \eqref{BLB2_opt_lemma12} and (\ref{neq1-lemma13})  implies that
\[
   Z(\pi) \;\ge\; \frac{1+\theta}{\theta} T_M + (1+\theta) p_u^A.
\]
Together with \eqref{delta_sigam_opt_lemma12}, this contradicts \eqref{neq2}.
\end{description}

Therefore, $J_u^A$ is largest job among those $A$-jobs arriving at or before time $t_u$.
\end{proof}

\begin{lemma}\label{lemma14}
If $r_{\min}(B)> t_u$, then $J_u^A$ starts at or after $t_u$ in $\pi$.
\end{lemma}
\begin{proof}
Suppose, for the sake of contradiction, that $J_u^A$ starts before $t_u$ in $\pi$. 
Since $r_{\min}(\mathcal{B})>t_u$, all jobs in $\mathcal{B}$ must start after $J_u^A$ in $\pi$. 
Consequently,
\begin{align}\label{lb1-opt-lemma15}
  C_{\max}^B(\pi)\;\ge\; p_u^A + P(\mathcal{B}'\cup\mathcal{B}).
\end{align}
    If  $J_u^A$ starts at or after $T_M$ in $\pi$, we further obtain that
\begin{align*}
  C_{\max}^B(\pi)\;\ge\; T_M+p_u^A + P(\mathcal{B}),
\end{align*}
    which, together with $C_{\max}^A(\pi)>C_{\max}^B(\pi)$, implies that
\begin{align*}
  Z(\pi)\;\ge\; (1+\theta)(T_M+p_u^A).
\end{align*}
Together with \eqref{delta_sigam_opt_lemma12}, this contradicts \eqref{neq2}.  
    Therefore, $J_u^A$ starts before $T_M$ in $\pi$, which implies that
     $J_u^A$ is released before $T_M$. By Observation \ref{observation1}, we have
\begin{align}\label{neq0_lemma14}
  p_u^A\geq \frac{\theta+1}{\theta}T_M.
\end{align}

 We consider two cases depending on whether $\overline{\mathcal{A}'}$ is empty.   
\begin{description}  
 \item[\bf{Case 1:}] $\overline{\mathcal{A}'}=\emptyset$. In this case, $P(\mathcal{A}')=p_u^A$.  
  Combining this with \eqref{z-sigma3}, \eqref{ALB_opt_lemma12}, and \eqref{lb1-opt-lemma15}, we obtain
  \begin{align*}
    Z(\sigma)-Z(\pi)\;\leq\;(1+\theta)T_M.
  \end{align*}
  Together with $C_{\max}^A(\pi)\geq p_u^A$, \eqref{lb1-opt-lemma15}, and \eqref{neq0_lemma14},  
this contradicts \eqref{neq2}.
   
\item[\bf{Case 2:}] $\overline{\mathcal{A}'}\neq\emptyset$.  
    Let $t$ be the starting time of the last job from $\overline{\mathcal{A}'}$ in $\sigma$.  
    By Lemma~\ref{lemma13}, $J_u^A$ is the largest job among those $A$-jobs released at or before $t$.  
    According to Procedure~A, we distinguish two subcases:  
 \begin{description}
   \item[\bf{Case 2.1:}] $T_M+P(\overline{\mathcal{A}'})\leq p_u^A$.  
  At time $t$, Procedure~A schedules the last job of $\overline{\mathcal{A}'}$ prior to $J_u^A$.  
  Hence,
   \[
 T_M+\frac{\theta^2}{(1+\theta)^2}P(\overline{\mathcal{A}'})\leq \frac{\theta}{1+\theta}p_u^A.
 \]
 
    Combining this with \eqref{z-sigma3}, \eqref{ALB_opt_lemma12},  \eqref{lb1-opt-lemma15} and $ P(\mathcal{A}')= P(\overline{\mathcal{A}'})+p_u^A$, we obtain
 \begin{align*}
   Z(\sigma)-Z(\pi)\leq \frac{\theta}{1+\theta}P(\mathcal{A}').
 \end{align*}
   Together with $Z(\pi)\geq P(\mathcal{A}')$, this contradicts \eqref{neq2}.

\item[\bf{Case 2.2:}] $T_M+P(\overline{\mathcal{A}'})> p_u^A$.  
  The last job of $\overline{\mathcal{A}'}$ is scheduled prior to the largest $A$-job $J_u^A$.  
  By Procedure~A, we have
\begin{align*}
   P(\overline{\mathcal{A}'})\geq \frac{\theta}{1+\theta}p_u^A+(\frac{1}{\theta}-\frac{\theta}{1+\theta})T_M.
\end{align*}  
   Combing this with $C_{max}^A(\pi)\geq P(\overline{\mathcal{A}'})+p_u^A$ and $C_{max}^B(\pi)> t_u\geq T_M+P(\overline{\mathcal{A}'})$, we deduce that
\begin{align*}
   Z(\pi)&\geq  \theta T_M  +p_u^A+(1+\theta)P(\overline{\mathcal{A}'})\\
         &\geq (1+\theta)p_u^A+\frac{1+\theta}{\theta}T_M.
\end{align*}
    This, together with \eqref{delta_sigam_opt_lemma12}, contradicts \eqref{neq2}. 
 \end{description}
\end{description}
This completes the proof. \qed 
\end{proof}

\begin{lemma}\label{lemma15}
If $r_{\min}(\mathcal{B})> t_u$, then $I$ is not a counterexample.
\end{lemma}

\begin{proof} 
By Lemmas~\ref{lemma10} and \ref{lemma11}, the condition $r_{\min}(\mathcal{B})>t_u$ implies that
$J_u^A\notin \mathcal{A}$ and $r_{\min}(\mathcal{A})\le t_v$.
The corresponding Gantt chart of $\sigma$ is shown in Fig.~\ref{Fig-sigma2}.

    If $r_{\min}(\mathcal{A}) > t_u$, then $J_u^A$ is the only available job at time $t_u$.
    By Algorithm AOLA, we have
\[
  t_u \;\ge\; \frac{\theta}{1+\theta}p_u^A.
\]
    
Moreover, by Lemma~\ref{lemma14}, the condition $r_{\min}(\mathcal{B})>t_u$ implies that $J_u^A$ starts at or after $t_u$ in $\pi$.
Combining this with $r_{\min}(\mathcal{A})>t_u$, $r_{\min}(\mathcal{B})>t_u$, and Lemma~\ref{lemma12}, we obtain
\[
  C_{\max}^A(\pi) \;\ge\; t_u + p_u^A + P(\mathcal{A}\cup \mathcal{B}).
\]
Combining these inequalities with \eqref{BLB2_opt_lemma12} and \eqref{z-sigma3}, we get
\[
  Z(\sigma) - Z(\pi) \;\le\; \frac{\theta}{1+\theta}\, Z(\pi).
\]
Therefore, $\mathcal{I}$ cannot be a counterexample.

In what follows, we assume $r_{\min}(\mathcal{A})\le t_u$.
Note that at time $t_u$ only $A$-jobs are available.
We distinguish two cases according to Procedure~A.
\begin{description}
  \item[\bf{Case 1:}] $T_M+P(\overline{\mathcal{A}'})+p_{s(t_u)}^A\leq p_u^A$.   
  At time $t_u$, $J_u^A$ is scheduled by Procedure A. Thus, 
  \begin{align*}
    T_M+\frac{\theta^2}{(1+\theta)^2}\big(P(\overline{\mathcal{A}'})+p_{s(t_u)}^A\big)>\frac{\theta}{1+\theta}p_u^A.
  \end{align*}
  Combining this with the condition of Case~1, we deduce that
  \begin{align}
    \label{neq1-case1-last-lemma}
    &T_M>\frac{\theta}{1+\theta}\big(P(\overline{\mathcal{A}'})+p_{s(t_u)}^A),\\
    \label{neq2-case1-last-lemma}
    &(1+\theta)T_M +P(\overline{\mathcal{A}'})+p_{s(t_u)}^A>\theta p_u^A.
  \end{align}
  From \eqref{neq1-case1-last-lemma} and Observation~\ref{observation1}, it follows that no jobs in $\overline{\mathcal{A}'}\cup \{J_{s(t_u)}^A\}$ are released before $T_M$. 
  Consequently, the job that can start before $T_M$ in $\pi$ is not in $\overline{A'}\cup \{J_{s(t_u)}^A\}\cup\{J_u^A\}$. 
  We further distinguish  three subcases.
\begin{description}
\item[\textbf{Case 1.1}:] In schedule $\pi$, there is one $A$-job that begins before $T_M$. 
  Then this $A$-job does not belong to $\overline{A'}\cup \{J_{s(t_u)}^A\}\cup\{J_u^A\}$ and all $B$-jobs will start after this $A$-job . 
  According to Observation \ref{observation1}, that $A$-job has a processing time of at least $\frac{1+\theta}{\theta}T_M$. Therefore, we have
  \begin{align}\label{CA_LB_lemma15}
  &C_{max}^A(\pi)\geq \frac{1+\theta}{\theta}T_M+P(\overline{A'})+p_u^A+p_{s(t_u)}^A,\\\label{CB_LB_lemma15}
  &C_{max}^B(\pi)\geq \frac{1+\theta}{\theta}T_M.
  \end{align}
    By combining these with \eqref{neq2-case1-last-lemma}, we obtain
 \begin{align*}
  Z(\pi)\geq  (1+\theta) p_u^A+\frac{1+\theta}{\theta}T_M,
  \end{align*}    
  which together with \eqref{delta_sigam_opt_lemma12}, implies that $I$ is not a counterexample.

\item[\textbf{Case 1.2}:]  In schedule $\pi$, suppose there is one $B$-job that begins before $T_M$. 
  Then all $A$-jobs will start after this $B$-job. 
  According to Observation \ref{observation1}, that $B$-job has a processing time of at least $(1+\theta)T_M$. Consequently, we have
  \begin{align*}
    &C_{max}^B(\pi)\geq (1+\theta)T_M,\\
    &C_{max}^A(\pi)\geq (1+\theta)T_M+P(\overline{A'})+p_u^A+p_{s(t_u)}^A.
  \end{align*}
  Combining these inequalities with \eqref{neq2-case1-last-lemma}, we obtain
  \begin{align*}
    Z(\pi) \;\ge\; \frac{\theta+1}{\theta} T_M + (1+\theta)p_u^A,
  \end{align*}
  which, together with \eqref{delta_sigam_opt_lemma12}, implies that $I$ is not a counterexample.

\item[\textbf{Case 1.3}:] No jobs start before $T_M$ in $\pi$. Then
\begin{align}\label{neq4-case1-last-lemma}
  C_{\max}^A(\pi) \;\ge\; T_M + P(\mathcal{B'}\cup \mathcal{B}) +p_u^A+ P(\overline{\mathcal{A}'})+P(\mathcal{A}).
\end{align}
This, together with \eqref{z-sigma3} and \eqref{BLB2_opt_lemma12}, yields
\begin{align}\label{neq6-case1-last-lemma}
  Z(\sigma)-Z(\pi) \;\le\; \theta p_u^A.
\end{align}
Combining this with \eqref{neq2-case1-last-lemma}, \eqref{neq4-case1-last-lemma}, $P(\mathcal{A})\geq p_{s(t_u)}^A$ and $C_{\max}^B(\pi)\ge T_M$, we conclude that $I$ is not a counterexample.

\end{description}   

  \item[\bf{Case 2}:] $T_M+P(\overline{\mathcal{A}'})+p_{s(t_u)}^A> p_u^A$.   
  At time $t_u$, Procedure A prioritizes the largest $A$-job, $J_u^A$, over all other available $A$-jobs, which results in the following inequality:
\begin{align*}
  P(\overline{\mathcal{A}'})+p_{s(t_u)}^A<\frac{\theta}{1+\theta}p_u^A+(\frac{1}{\theta}-\frac{\theta}{1+\theta})T_M.
\end{align*} 
  Substituting this bound into the initial condition, we obtain that
\begin{align}\label{neq2-case2-last-lemma}
  &T_M>\frac{\theta}{2\theta+1}p_u^A,\\\label{neq3-case2-last-lemma}
  &T_M >\frac{\theta}{1+\theta}(P(\overline{\mathcal{A}'})+p_{s(t_u)}^A)).
\end{align}  
   
   Based on Observation \ref{observation1}, the last inequality implies that no jobs in the set $\overline{A'}\cup \{J_{s(t_u)}^A\}$ are released before $T_M$. 
   Consequently, the job that can start before $T_M$ in $\pi$ is not in $\overline{A'}\cup \{J_{s(t_u)}^A\}\cup\{J_u^A\}$. We then consider three subcases.
\begin{description}
\item[\textbf{Case 2.1}:] In schedule $\pi$, there is one $A$-job that begins before $T_M$. 
  As in Case 1.1, \eqref{CA_LB_lemma15} and \eqref{CB_LB_lemma15} hold.
Combining these with the initial assumption of Case 2,  \eqref{neq2-case2-last-lemma},
and the condition $1+\theta\ge\theta^2$, we obtain
\[
  Z(\pi)\ge (1+\theta)p_u^A + \frac{1+\theta}{\theta}T_M.
\]
Together with \eqref{delta_sigam_opt_lemma12}, this shows that $I$ is not a counterexample.
 
\item[\textbf{Case 2.2}:] In schedule $\pi$, there is one $B$-job that begins before $T_M$. 
  Then all $A$-jobs will start after this $B$-job. 
  According to Observation \ref{observation1}, that $B$-job has a processing time of at least $(1+\theta)T_M$. Therefore, we have
  \begin{align*}  
  C_{max}^A(\pi)\geq (1+\theta)T_M+P(\overline{A'})+p_u^A+p_{s(t_u)}^A,
  \end{align*}
  which, combined with the initial condition of Case 2, \eqref{neq2-case2-last-lemma} and  the condition $\theta+1\geq \theta^2$, implies that
 \begin{align}
  C_{max}^A(\pi)\geq  (1+\theta) p_u^A.
  \end{align}    
  Combining this result with $C_{max}^B(\pi)\geq (1+\theta)T_M$ and \eqref{delta_sigam_opt_lemma12}, we can conclude that $I$ is not a counterexample.

\item[\textbf{Case 2.3}:] No jobs start before $T_M$ in $\pi$. Thus, 
\begin{align*}
    C_{\max}^A(\pi)\geq T_M+P(\overline{A'})+p_u^A+p_{s(t_u)}^A,
\end{align*}
  which, combined with the initial condition of Case 2, yields that $C_{\max}^A(\pi)\geq 2p_u^A$. 
  Together with $C_{\max}^B(\pi)\ge T_M$, \eqref{neq2-case2-last-lemma}, and the condition $\theta+1\ge \theta^2$, this implies that
\begin{align*}
  Z(\pi)\;\ge\; (1+\theta)\,p_u^A.
\end{align*}

Following a similar argument to Case 1.3, it can be shown that \eqref{neq6-case1-last-lemma} holds. Therefore, $I$ is not a counterexample in this subcase.  

\end{description}   
\end{description}
This completes the proof. \qed
\end{proof}

    Analogously to Lemma~\ref{lemma15}, we can show that $\mathcal{I}$ is not a counterexample when $r_{\min}(\mathcal{A})>t_v$.
However, by Lemmas~\ref{lemma10} and \ref{lemma11}, the conditions $r_{\min}(\mathcal{B})>t_u$ and $r_{\min}(\mathcal{A})>t_v$ cannot occur simultaneously.
Therefore, no counterexample exists. We thus obtain the following theorem.
\begin{theorem}
   Algorithm has a competitive ratio of $1+\frac{\theta}{1+\theta}$ when $1\leq \theta \leq \frac{\sqrt{5}+1}{2}$.
\end{theorem}

\section{Conclusions}
\label{sec:8}

This paper studied a two-agent single-machine scheduling problem 
$1\mid r_j, \text{online}\mid C_{\max}^A+\theta C_{\max}^B$, where 
$1 \leq \theta \leq \tfrac{1+\sqrt{5}}{2}$. 
For this problem, we proposed Algorithm ALOA, which achieves the best possible 
competitive ratio of $1+\tfrac{\theta}{1+\theta}$. 
This improves upon the result of Zhao et al., who presented an online algorithm 
with a competitive ratio of $\tfrac{1+2\theta+\theta^2}{2+2\theta+\theta^2}$.

When $\theta > \tfrac{1+\sqrt{5}}{2}$, it becomes more advantageous to process 
$B$-jobs before $A$-jobs. 
However, it is difficult to make scheduling decisions when only $A$-jobs are 
available and $t < \tfrac{\theta}{1+\theta}p_{l(t)}^A$. 
From the analysis of Algorithm ALOA, we find that only in Case~2 of 
Lemma~\ref{lemma15}, ensuring the competitive ratio of 
$1+\tfrac{\theta}{1+\theta}$ requires that $1+\theta \geq \theta^2$, 
that is, $\theta \leq \tfrac{1+\sqrt{5}}{2}$. 
At a time $t$ with $t < \tfrac{\theta}{1+\theta}p_{l(t)}^A$, 
regardless of whether the algorithm schedules $J_{l(t)}$ or $J_{s(t)}$, 
the competitive ratio---the objective function value of the algorithm’s solution 
relative to the optimal one---may exceed $1+\tfrac{\theta}{1+\theta}$. 
To guarantee this bound, it is necessary to introduce an appropriate waiting 
strategy when scheduling jobs in $\mathcal{A}'$. 
Designing such a strategy remains a key challenge. 
In particular, when $\theta > \tfrac{1+\sqrt{5}}{2}$, 
the current algorithm alone cannot ensure the performance ratio 
$1+\tfrac{\theta}{1+\theta}$.

An immediate direction for future research is to develop an online algorithm that can achieve a competitive ratio of $1+\tfrac{\theta}{1+\theta}$ for the case where $\theta > \tfrac{1+\sqrt{5}}{2}$. 
This would require a sophisticated waiting strategy or alternative scheduling approach. It would also be valuable to explore how these results extend to multi-machine environments or other objective functions.
\section*{Acknowledgements}
This research is supported by National Natural Science of China (12371317, 12571338).




\end{document}